\documentclass{hep99}
\usepackage{epsf}
\newcommand{\be}{\begin{equation}}
\newcommand{\ee}{\end{equation}}
\newcommand{\bea}{\begin{eqnarray}}
\newcommand{\eea}{\end{eqnarray}}
\newcommand{\bean}{\begin{eqnarray*}}
\newcommand{\eean}{\end{eqnarray*}}

\newcommand{\AmS}{{\protect\the\textfont2
  A\kern-.1667em\lower.5ex\hbox{M}\kern-.125emS}}

\begin{document}

% declarations for front matter
\title{Exotic Quarkonia from Lattice QCD}

\author{T. Manke for the CP-PACS Collaboration 
\footnote{CP-PACS Collaboration: A.~Ali~Khan, S.~Aoki,
R.~Burkhalter, S.~Ejiri, M.~Fukugita, S.~Hashimoto, N.~Ishizuka,  Y.~Iwasaki,
K.~Kanaya, T.~Kaneko, Y.~Kuramashi, T.~M., K.~Nagai, M.~Okawa, H.P.~Shanahan,
A.~Ukawa  and T.~Yoshi\'e}
}

\address{Center for Computational Physics, University of Tsukuba, Tsukuba, Ibaraki 305, Japan}
       
\abstract{
We present non-perturbative results for the spectrum of heavy quarkonia. 
Using an anisotropic formulation of Lattice QCD we achieved an unprecedented
control over statistical and systematic errors. We also study relativistic
corrections to the leading order predictions for heavy hybrids and
conventional bound states.}
\maketitle
\section{INTRODUCTION 
%\footnote{CP-PACS Collaboration: A.~Ali~Khan, S.~Aoki,
%R.~Burkhalter, S.~Ejiri, M.~Fukugita, S.~Hashimoto, N.~Ishizuka,  Y.~Iwasaki,
%K.~Kanaya, T.~Kaneko, Y.~Kuramashi, T.~M., K.~Nagai, M.~Okawa, H.P.~Shanahan,
%A.~Ukawa  and T.~Yoshi\'e}
}
Lattice studies play an important role for our theoretical understanding of QCD.
However, the conventional approach is not well suited to accommodate physical systems
with widely separate scales. When studying heavy quarkonia on {\it isotropic} lattices several approximations
have to be made to render the numerical simulations tractable. In this context
a non-relativistic approach (NRQCD) has lead to very precise calculations
of the low lying spectrum in heavy quarkonia
\cite{NRQCD_omv4,NRQCD_omv6}. 
New complications arise if high energetic excitations are to be resolved, for
which the temporal discretisation is often too coarse and the correlator 
of such heavy states cannot be measured accurately for long times \cite{cm_hybrid,milc_hyb,ukqcd_hyb}.
More recently {\it anisotropic} lattices were demonstrated to circumvent 
this problem by giving the lattice a fine temporal resolution whilst
maintaining a coarse discretisation in the spatial direction \cite{morning_glue98}. This approach has resulted 
in very encouraging results for the spectrum of glueballs and hybrid states,
which are of particular interest as they are non-perturbative revelations of
the gluon degrees of freedom \cite{morning_glue98,cppacs_hybrid98,kuti_hybrid98}.
Here we extend those methods to investigate
also other excitations and the spin structure in heavy quarkonia 
more carefully.

\section{RESULTS}
\begin{figure}[htb]
\hbox{\epsfxsize = 80mm  \epsfysize = 70mm \epsffile{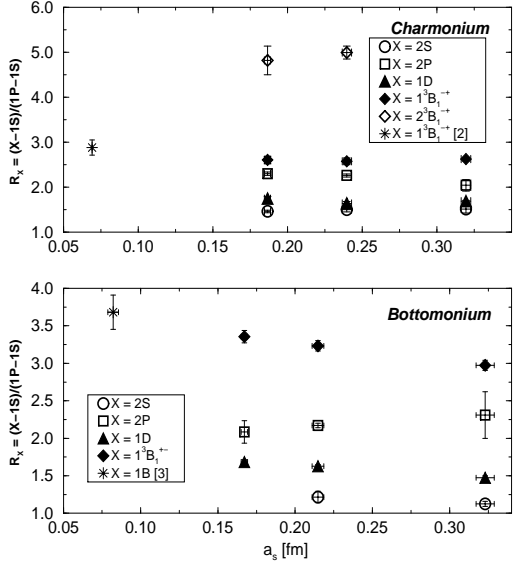}}
%\vskip -5pt
\caption{Scaling analysis for excited Quarkonia. We plot the ratio $R_X$
against the spatial lattice spacing for different states X = S,P,D
and ${^3}B_1^{-+}$ (exotic hybrid).}
\label{fig:RX}
%\vskip -20pt
\end{figure}

In order to study excited quarkonia with small statistical errors 
we employ an anisotropic and spatially coarse gluon action and the NRQCD
approach for the forward propagation of the heavy quarks.
The latter is improved to contain spin-dependent terms up to ${\cal O}(mv^6)$.
From the implementation details given in \cite{qcd99} we expect discretisation 
errors of ${\cal O}(a_s^4,a_t^2)$. Radiative corrections to this classical
result are reduced by imposing a mean-field improvement on all gauge links.
In Fig. \ref{fig:RX} we present our results for the excitations in Charmonium
and Bottomonium as normalised to the $1P-1S$ splitting. In each case
we have studied bound states with several different quantum numbers, including the exotic hybrid, ${^3}B_1^{-+}$.
The possibility to resolve all these states reliably
is clearly due to the fine temporal resolution of our lattices.
Using statistical ensembles of up to a few thousand independent measurements, we could rely
on very simple-minded hadron operators to obtain the excitation energies from correlated multi-exponential fits.
 
Within the NRQCD approach it is paramount to establish a scaling region
for physical quantities already at finite lattice spacing. Our analysis demonstrates
the existence of such a window for the excitations in Charmonium and
Bottomonium \cite{cppacs_hybrid98}.  Moreover, the excellent agreement of the lowest lying charmonium hybrid with a 
relativistic calculation on isotropic lattices should be considered a
combined success of anisotropic lattices and the NRQCD approach. 
Owing to the efficiency of this new approach we were able to test our results
against finite size effects. For the lowest lying $c\bar cg$ hybrid we could not resolve
any change for volumes larger than 1.2 fm and we presently continue this analysis for all other excitations.
For the Bottomonium hybrid we have also consistent results from two different
anisotropies, which confirms our initial assumption of small temporal lattice
spacings artefacts.

\begin{figure}[t]
\hbox{\epsfxsize = 80mm  \epsfysize = 70mm \epsffile{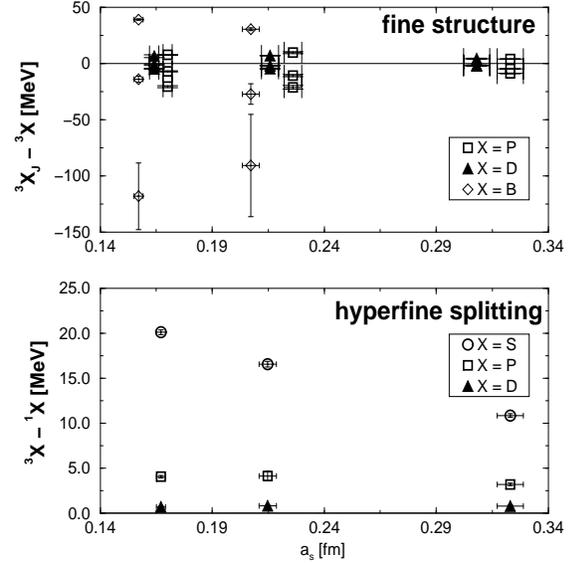}}
%\vskip -5pt
\caption{Spin structure in Bottomonium for states with different orbital
momentum and magnetic hybrids, diamonds.}  
\label{fig:spin}
%\vskip -20pt
\end{figure}

The inclusion of relativistic corrections is a significant improvement over
previous NRQCD calculations of hybrid states, which were restricted to only
leading order in the velocity expansion. At this level there are no spin-dependent operators
and we have a strict degeneracy of all singlet and triplet states.
Here we also include spin terms to break this degeneracy.
In particular, we could directly observe the exotic hybrid,
which is the state of greatest phenomenological interest.
Our results for the spin structure in Bottomonium are shown in Fig. \ref{fig:spin}. Similar results for
Charmonium are presented elsewhere \cite{qcd99}.
We notice a clear reduction of the fine structure in D-states when
compared to that of P-states. Similarly, the hyperfine
splitting, ${^3}X - {^1}X$, is suppressed as the orbital angular momentum is increased. 
This is in accordance with expectations from potential models.
Our data also indicates that the fine structure in hybrid states 
is enlarged as the result of the gluon angular momentum to which the spin can
couple. 

Finally one should notice that the ${^3}S_1-{^1}S_0$ splitting does not yet scale 
on the lattices considered here. It is apparent that one needs a better improvement description to
account for lattice spacing artefacts in such UV-sensitive quantities.

In conclusion, we find that coarse and anisotropic lattices
are extremely useful for precision measurements of higher excited states.
This is due to an improved resolution in the temporal direction
and the possibility to generate large ensembles of gauge field configurations 
at small computational cost.
It has lead to an unprecedented control over statistical and systematic
errors in lattice studies of heavy quarkonia. 
We could observe a clear hierarchy in the spin structure, depending on the orbital angular momentum. 
The remaining systematic error for all our predictions is an uncertainty in
the scale as the result of the quenched approximation. 
This is not yet controlled and we find a variation of 10-20\%,
depending on which experimental quantity is used to set the scale.

This work is supported by the Research for the Future Programme of the JSPS.


\begin{thebibliography}{9}
%\bibitem{NRQCD} G.P. Lepage {\it et al.} Phys.Rev.{\bf D}46 (1992) 4052. 
%\bibitem{FERMILAB} A.X. El-Khadra {\it et al.},  Phys.Rev.{\bf D}55 (1997) 3933. 
\bibitem{NRQCD_omv4} G.P. Lepage {\it et al.} Phys.Rev.{\bf D}46 (1992) 4052; C.T.H. Davies {\it et al.}, Phys.Rev.{\bf D}50
(1994) 6963; 
\bibitem{NRQCD_omv6} H. Trottier, Phys.Rev.{\bf D}55 (1997) 6844;\\
T. Manke {\it et al.}, Phys.Lett.{\bf B}408 (1997) 308;\\
N. Eicker {\it et al.}, Phys.Rev. {\bf D}57 (1998), 4080. 
\bibitem{cm_hybrid} N.A. Campbell {\it et al.} Phys.Lett.{\bf 142} (1984) 291.
\bibitem{milc_hyb} C. Bernard {\it et al.}, Phys.Rev.{\bf D}56 (1997) 7039.
\bibitem{ukqcd_hyb} T. Manke {\it et al.}, Phys.Rev.{\bf D}57 (1998) 3829.
\bibitem{morning_glue98} C.J. Morningstar and M. Peardon, Phys.Rev. {\bf D}56
 (1997) 4043.
\bibitem{cppacs_hybrid98} T. Manke {\it et al.} (CP-PACS Collaboration), Phys.Rev.Lett.82 (1999) 4396.
\bibitem{kuti_hybrid98} K.J. Juge {\it et al.} Phys.Rev.Lett.82 (1999) 4400.
\bibitem{qcd99} T. Manke, hep-lat/9909038, talk at QCD'99, to appear in Nucl.Phys.(Proc.Suppl.).
\end{thebibliography}
\end{document}